\documentclass[prl,twocolumn,aps,amssymb,amsmath,superscriptaddress,showpacs]{revtex4-1}

\usepackage{bm}
\usepackage{graphicx}
\usepackage{color}
\usepackage{mathtools}
\usepackage{amsfonts}
\usepackage{textcomp} 
\usepackage{microtype} 

\usepackage{hyperref}
\hypersetup{colorlinks=true}
\usepackage[all]{hypcap} 

\newcommand{\w}{\omega}

\newcommand{\mr}[1]{\mathrm{#1}}

\begin{document}

\title{Fate of spinons at the Mott point}

\author{Tsung-Han Lee}
\affiliation{Department of Physics and National High Magnetic Field Laboratory, Florida
State University, Tallahassee, FL 32310}
\author{Serge Florens}
\affiliation{Institut N\'{e}el, CNRS and Universit\'e Grenoble Alpes, F-38042 Grenoble, France}
\author{Vladimir Dobrosavljevi\'c}
\affiliation{Department of Physics and National High Magnetic Field Laboratory, Florida
State University, Tallahassee, FL 32310}

\begin{abstract}
Gapless spin liquids have recently been observed in several frustrated Mott
insulators, with elementary spin excitations - ``spinons" - reminiscent of
degenerate Fermi systems. However, their precise role at the Mott point, where
charge fluctuations begin to proliferate, remains controversial and
ill-understood. Here we present the simplest theoretical framework that treats
the dynamics of emergent spin and charge excitations on the same footing, 
providing a new physical picture of the Mott metal-insulator transition at half filing. 
We identify a generic orthogonality mechanism leading to strong damping of spinons, 
arising as soon as the Mott gap closes. Our results indicates that spinons should not play a
significant role within the high-temperature quantum critical regime above the
Mott point - in striking agreement with all available experiments. 
\end{abstract}

\date{\today}

\maketitle

{\it Introduction.}
The physical nature of the Mott metal to insulator transition (MIT), a
phenomenon generic to strongly correlated materials, still remains the subject
of much controversy and debate. In contrast to conventional critical phenomena,
the relevant degrees of freedom at the MIT cannot be easily identified using an
appropriate order parameter of symmetry breaking principle, although different
competing orders often do arise in its vicinity. As stressed in pioneering works
by Mott and Anderson, however, a fundamentally different physical mechanism has
to exist, because the Mott insulating state typically persists to temperatures
much higher than any conventional order. 

A clear physical picture of how a sharp Mott transition can exist without any
intervening order has emerged only recently, with the development of Dynamically
Mean-Field Theory (DMFT)~\cite{DMFTRevModPhys.68.13}, which is formally exact in the
limit of large coordination. Physically, it represents the limit of maximal
frustration, therefore eliminating the precursors of any competing order, and
retaining only purely local dynamical scattering processes. In contrast to
alternative theoretical approaches describing dilute low-energy excitations,
DMFT is most reliable at intermediate and high temperatures, where incoherent
behavior prevails. The DMFT picture resulted in a finite temperature
first-order boundary between the two phases, around which many fascinating
phenomena organize themselves. Several of its most striking predictions were
experimentally confirmed in a variety of systems, including organic
charge-transfer salts of the $\kappa$-family, as well as various transition
metal oxides. Here one can list the observation of strongly renormalized Fermi
liquids \cite{AGeorges_organic_exp_prl}, bad metal behavior, Ising universality
near the Mott endpoint~\cite{Limelette,Kagawa}, and even quantum critical
scaling at higher temperatures~\cite{VladMottCritical,KanodaMottCriticalExp}.

At lower temperatures, most Mott insulators still undergo magnetic ordering, but
experimental studies of several frustrated organic materials have instead
observed spin liquid behavior
\cite{YKurosakiBEDT,MYamashitagaped,SYamashitagapless,SYamashitagapless2011,MYamashita04062010gapless}.
Two particular compounds have attracted a lot of attention,
$\kappa$-(BEDT-TTF)$_2$Cu$_2$(CN)$_3$ and EtMe$_3$Sb[Pd(dmit)$_2$]$_2$. Most
remarkably, thermodynamic measurement in the Mott insulating phase of these
materials have revealed behavior normally expected for metals, including
displaying linear in temperature specific heat and large thermal conductivity,
indicating the presence of gapless magnetic excitations. 

\begin{figure}[t]
\includegraphics[width=0.7\linewidth,height=0.28\linewidth]{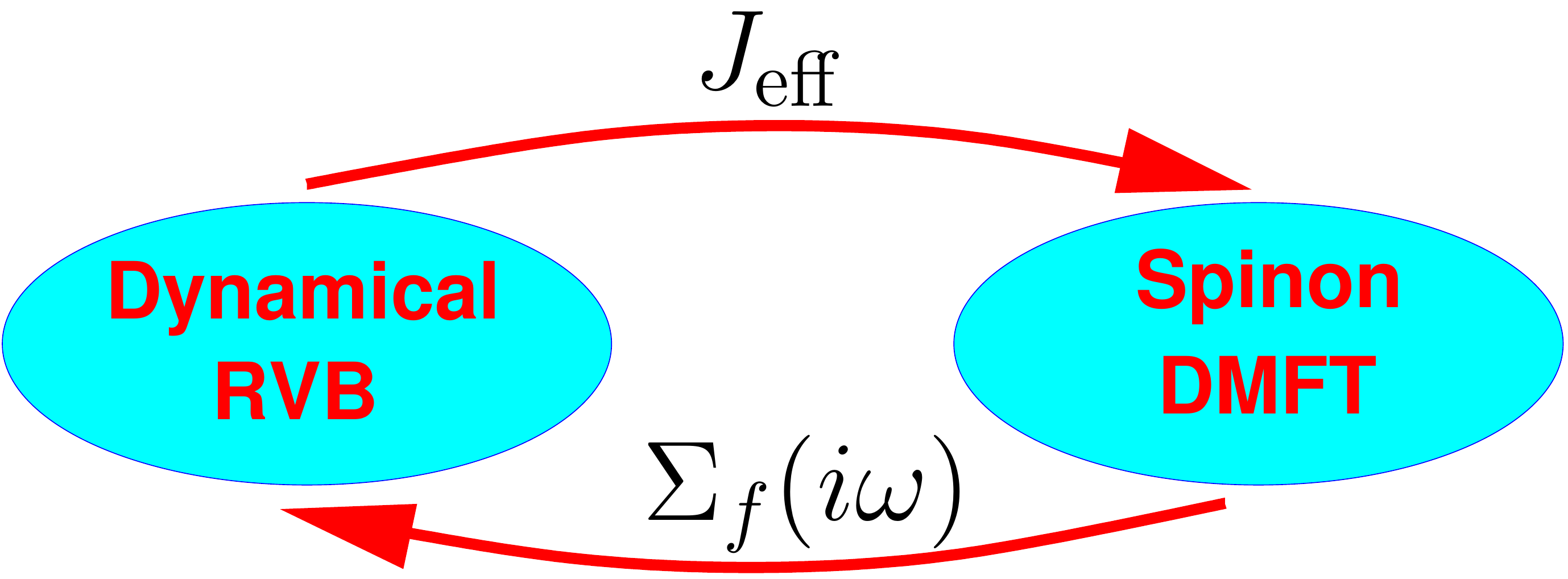}
\caption{(Color online) Proposed scheme to blend the static RVB approach with
the DMFT, using an impurity solver based on emergent spinon degrees of freedom.
The spinon self-energy $\Sigma_f(i\w)$, usually absent from a mean-field RVB 
treatment, is extracted from a spinon-based DMFT framework. 
Using the RVB self-consistent determination of the spinon bandwidth
$J_\mathrm{eff}$, this in turn allows to incorporate feedback effects 
due to scattering on charge excitations, which affect the gapless 
spin-excitations in the Mott state.}
\label{scheme}
\end{figure}

These observations are easiest to rationalize using a time-honored idea, the 
resonating valence bond (RVB) theory~\cite{SlaveBosonPatrickLeeRevModPhys}, 
in which chargeless spin excitations embody a magnetic fluid with fermionic statistics.
Despite this appealing picture, the RVB approach focuses on the zero temperature
phases, and is at trouble in recovering the expected DMFT predictions upon warming 
up the system. One question that immediately arises concerns the compatibility
of the two electronic fluids, namely the Fermi liquid describing the correlated
metallic state, and the magnetic fluid characterizing such a gapless Mott insulator.
In this Letter, we address this question by proposing a consistent framework that 
succeeds in marrying the DMFT and the RVB approaches. This allows to preserve the
known high temperature phenomenology of the DMFT, while introducing the strong 
thermodynamic signature of a gapless spin liquid state in the Mott phase. 
Akin to the impossibility of mixing oil and vinegar, we find that the Fermi liquid 
and the magnetic fluid are not miscible into each other in the transition region. 
While the insulating spin-liquid obviously cannot support mobile charge carriers,
the Fermi liquid metal always shows strongly incoherent spinon excitations, due
to an orthogonality catastrophe~\cite{Costi_NCA1,Costi_NCA2,Anderson_Orthogonality}.
This observation has important consequences, because the two associated Fermi surfaces 
cannot be tuned into each other in a continuous way.
Thus, the Mott localization into a gapless spin-liquid turns out to
have first-order character even at zero temperature, similarly to the case of a 
spin-gapped insulator described by frozen short-range singlets. This result is in
contrast with other scenarios based on the RVB picture only, in which the quasiparticles vanish 
continuously~\cite{sergeprb1_SR_Meanfield,SSLeePatrickLeePRL,SenthilKimSpinliquidPRL,SenthilCriticalFS}, 
but is consistent with all available experiments. 

We start by describing our theoretical framework, that is sketched in
Fig.~\ref{scheme}. 
The main idea is to solve the DMFT equations by explicitly
introducing spinon degrees of freedom, and to feedback the resulting spinon
self-energy $\Sigma_f(\omega)$ into the RVB equations, thus affecting the
stability of the spin liquid.
To be concrete, we consider henceforth the half-filled Hubbard-Heisenberg Hamiltonian:
\begin{eqnarray}
\nonumber
H &=& -t\sum_{\left<i,j\right>\sigma}d_{i\sigma}^{\dagger}d_{j\sigma}
+U\sum_{i}\left[d_{i\uparrow}^{\dagger}d_{i\uparrow}-\frac{1}{2}\right]
\left[d_{i\downarrow}^{\dagger}d_{i\downarrow}-\frac{1}{2}\right]\\
&& + J\sum_{\left<i,j\right>}
\Bigg[\sum_{\alpha,\alpha'} d_{i\alpha}^{\dagger}
\frac{\vec{\tau}_{\alpha,\alpha'}}{2}d_{i\alpha'}\Bigg]
. \Bigg[\sum_{\sigma,\sigma'} d_{j\sigma}^{\dagger}
\frac{\vec{\tau}_{\sigma,\sigma'}}{2}d_{j\sigma'}\Bigg].
\label{eq:ham}
\end{eqnarray}
Here $t$ is the intersite hopping, $U$ the local Coulomb interaction, $J$ an
explicit nearest neighbor exchange, and we have denoted by $\vec\tau$ the set
of Pauli matrices.
The RVB picture results here from a decomposition of the physical electron
into a chargeless spin-carrying fermion $f^\dagger_{j,\sigma}$ and a spinless
charge-carrying compact boson $X_j=e^{i\theta_j}$, using on each site $j$ 
the so-called slave rotor~\cite{sergeprb1_SR_Meanfield,sergeprb2_imp_solver} 
decomposition $d^\dagger_{j,\sigma} =
f^\dagger_{j,\sigma} e^{i\theta_j}$, with $i\partial/\partial \theta_j = 
\sum_\sigma[f_{j\sigma}^{\dagger}f_{j\sigma}-\frac{1}{2}]$. This provides an 
effective Hamiltonian:
\begin{eqnarray}
H_\mathrm{eff}=
\sum_{\left<i,j\right>\sigma}f_{i\sigma}^{\dagger}f_{j\sigma}
\left[J_\mathrm{eff}-t e^{i(\theta_{i}-\theta_{j})}\right]
-\frac{U}{2}\sum_{i}\frac{\partial^2}{\partial \theta_i^2},
\label{eq:ham-rvb}
\end{eqnarray}
with $J_\mathrm{eff} = - J \big<f_{i\sigma}^{\dagger}f_{j\sigma}\big>$
the RVB bond parameter. This approximation of the Heisenberg term
assumes the formation of a spinon Fermi surface, yet neglects a possible
momentum-dependent spinon self-energy. Such non-local effects, associated
to fluctuations beyond the RVB mean-field, are not expected to change
the physics of the spin-liquid Mott insulator, owing to the strong stability 
of such zero-entropy charge-gapped state.
The mean-field Hamiltonian~(\ref{eq:ham-rvb}) is consecutively solved within the DMFT, by
employing an impurity solver that is naturally based on the spinon/rotor
decomposition~\cite{sergeprb2_imp_solver}, leading to the following respective 
local Green's functions in Matsubara domain:
\begin{eqnarray}
G_{f}(i\omega_{n})^{-1}&=&i\omega_{n}+\mu-\Sigma_{f}(i\omega_{n})-\Delta_{f}(i\omega_{n}),\\
G_{X}(i\nu_{n})^{-1}&=&\frac{\nu_{n}^{2}}{U}+\lambda-\Sigma_{X}(i\nu_{n}),
\end{eqnarray}
and self-energies in imaginary time:
\begin{eqnarray}
\Sigma_{f}(\tau)&=&\Delta(\tau)G_{X}(\tau),\\
\Sigma_{X}(\tau)&=& \mathcal{N}\Delta(\tau)G_{f}(\tau),
\end{eqnarray}
with $\mathcal{N}=3$ to ensure the right shape of the Mott phase diagram, 
and $\lambda$ a Lagrange multiplier that enforces the constraint $|X_j|^2=1$ in 
average, namely $G_X(\tau=0)=1$.
The DMFT self-consistency equations account both for the physical electron
and the spinon hybridization functions, which read for the Bethe lattice: 
$\Delta(\tau)=t^{2}G_{d}(\tau)=t^2G_f(\tau)G_X(\tau)$ and
$\Delta_{f}(\tau)=J_\mathrm{eff}^{2}G_{f}(\tau)$.
The generic RVB equation, $J_\mathrm{eff} = - J G_{f}(\left<i,j\right>,\tau=0)$ can 
now be expressed on the Bethe lattice:
\begin{equation}
J_\mathrm{eff}=\frac{J}{2\beta}\underset{i\omega_{n}}{\sum}
\frac{J_\mathrm{eff}}{\frac{z_n^{2}}{2}
+z_n \sqrt{\frac{z_n^{2}}{4}
+J_\mathrm{eff}^{2}}+J_\mathrm{eff}^{2}},
\label{eq:MF_eq_correction}
\end{equation}
with $z_n = \omega_{n}-\mathrm{Im}\Sigma_{f}(i\omega_{n})$ and $\beta=1/T$ the
inverse temperature. We will set in what follows $t=1/2$, taking the
electronic half-bandwith $D=2t=1$ as natural energy unit.
The computation of the electronic specific heat results from the internal
energy per site (with $N_s$ sites):
\begin{eqnarray}
\nonumber
\frac{\big<H\big>}{N_s} &=& \frac{2}{\beta N_s} 
\sum_{n,k}[\epsilon_k^d G_d(k,i\w_n)
+\epsilon_k^f G_f(k,i\w_n)] e^{i\w_n 0^+}\\
&&
+ \frac{U}{2} D_{\uparrow\downarrow}
+ \frac{1}{N_s}  \sum_{\left<i,j\right>\sigma}
\frac{J_\mathrm{eff}^2}{J} ,
\label{Internal}
\end{eqnarray}
with $\epsilon_k^{d/f}$ the electron and spinon dispersion relations, 
and $G_{d/f}(k,i\w_n)$ their respective lattice Green's functions.
For the Bethe lattice, the sum over momentum in Eq.~(\ref{Internal})
can be replaced by an energy integral over the corresponding semi-circular
density of states~\cite{SupInfo}.
An important term to consider here is the double occupancy 
$D_{\uparrow\downarrow}$, which 
is related to the local charge susceptibility by 
$D_{\uparrow\downarrow}=(1/2)\chi_c(\tau=0)$. 
This quantity can be expressed from either a spinon response $\chi_c^f$ or a 
rotor response $\chi_c^X$:
\begin{eqnarray}
\nonumber
\chi_c^f(\tau) &=& 
\big<\sum_{\sigma,\sigma'}[f_{j\sigma}^{\dagger}(\tau)f_{j\sigma}(\tau)-\frac{1}{2}]
[f_{j\sigma'}^{\dagger}(0)f_{j\sigma'}(0)-\frac{1}{2}] \big>, \\
\chi_c^X(\tau) &=& \big<i\frac{\partial}{\partial \theta_j}(\tau) 
i\frac{\partial}{\partial \theta_j}(0)\big>.
\end{eqnarray}
Both expressions are equivalent only provided the constraint is dealt strictly,
but in a mean-field treatment, one must use Nagaosa and Lee's composition
rule~\cite{SlaveBosonPatrickLeeRevModPhys,SupInfo}, 
$\chi_c(i\w) = [(\chi_c^f)^{-1}+(\chi_c^X)^{-1}]^{-1}$.

The solution of the combined RVB and DMFT self-consistent scheme provides
the physical density of states and specific heat curves shown in
Fig.~\ref{rhoCv}.
\begin{figure}[th]
\includegraphics[width=0.99\linewidth,height=0.5\linewidth]{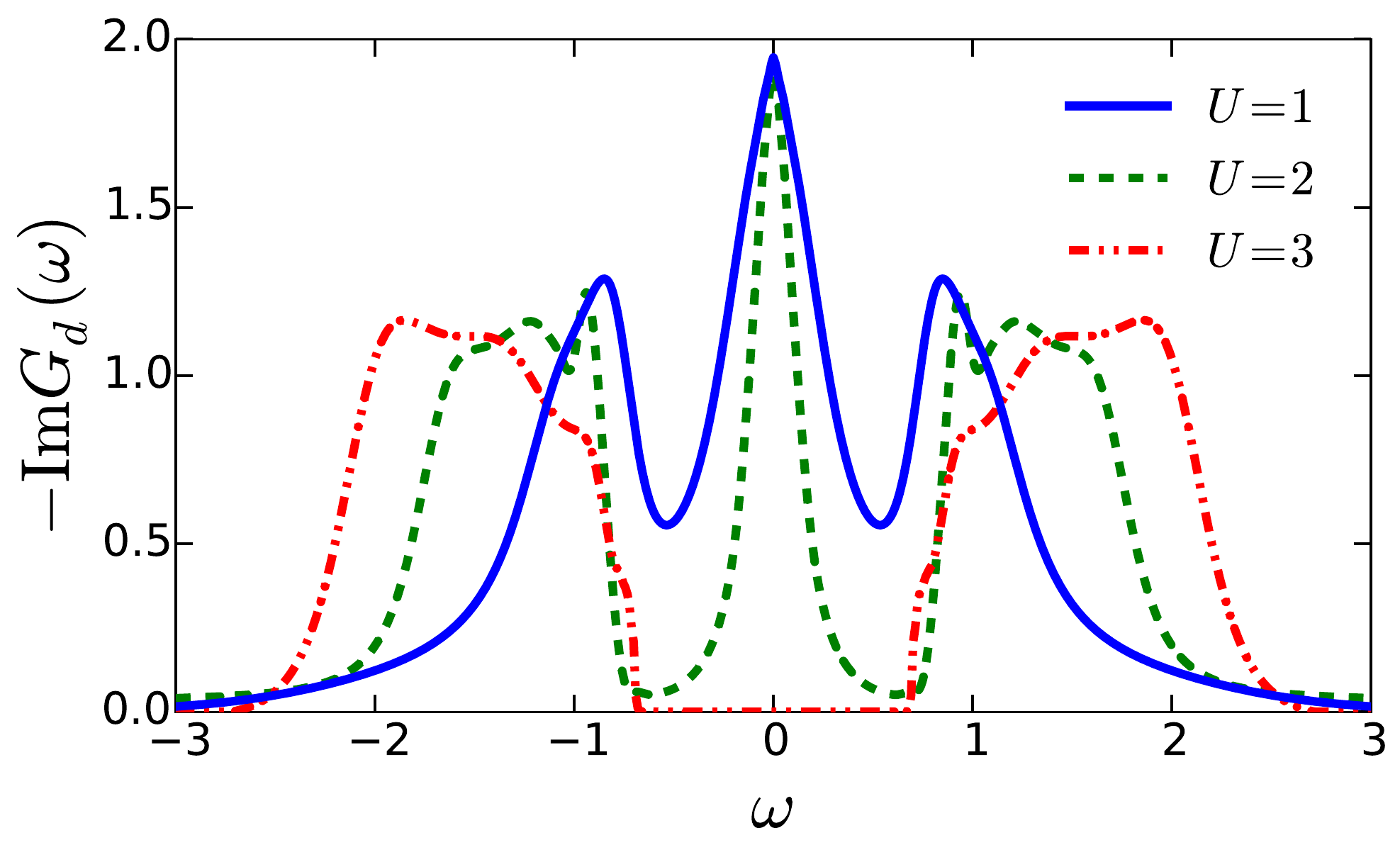}
\vspace{-0.6cm}\\
\includegraphics[width=0.99\linewidth,height=0.5\linewidth]{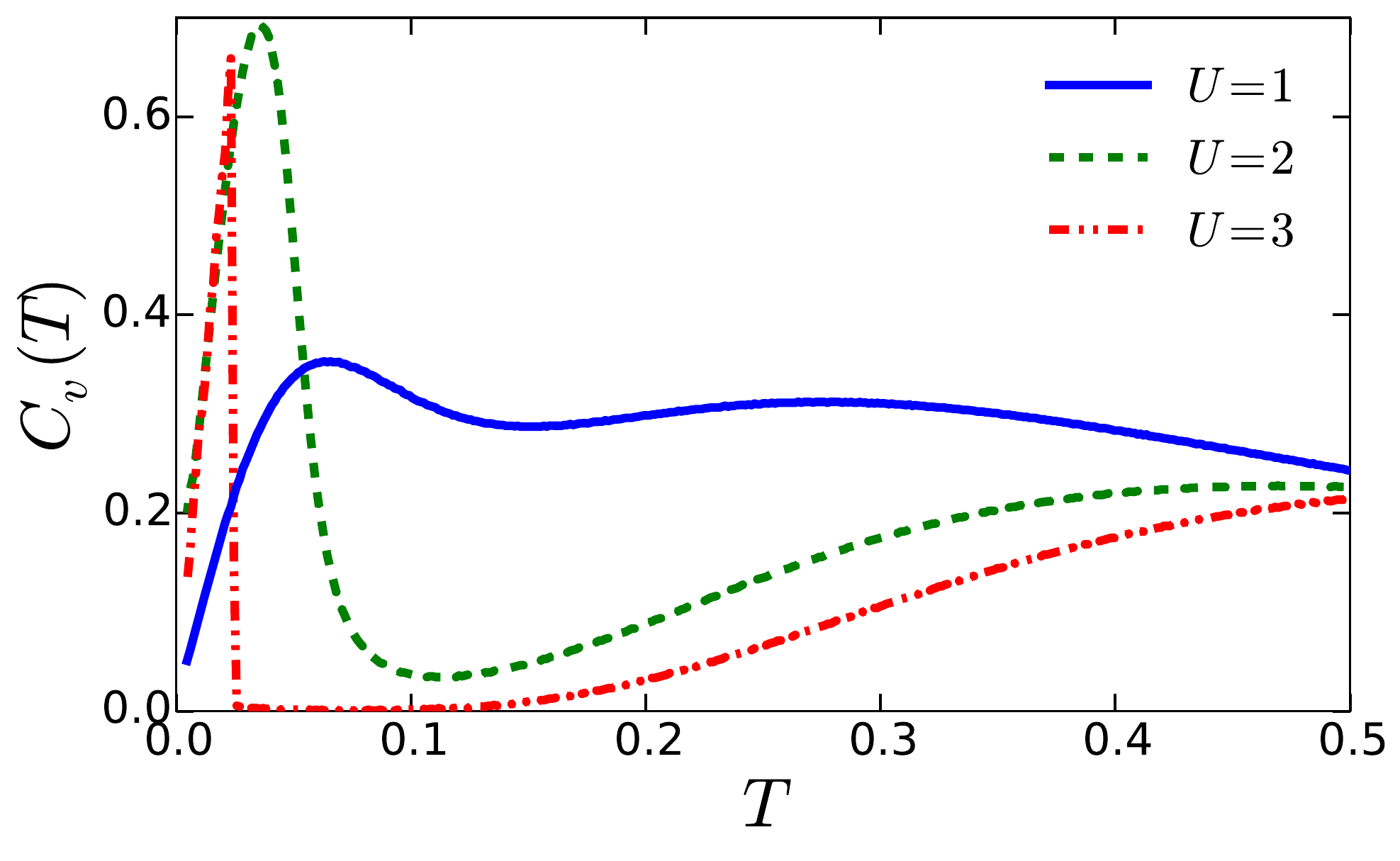}
\caption{(Color online) Upper panel: electronic density of states across
the Mott transition for $T=0.005D$ and $J/D=0.2$, with increasing values of the
Coulomb interaction $U/D=1,2,3$. Lower panel: corresponding specific heat
as a function of temperature. A striking linear in temperature contribution 
remains in $C_V$ at $U/D=3$, while quasiparticles have disappeared from the 
density of states in the Mott phase.}
\label{rhoCv}
\end{figure}
The electronic density of states shows at low temperature the expected behavior:
the quasiparticle peak narrows down for increasing values of $U$, with strong
spectral weight transfer towards Hubbard bands located at $\pm U/2$. This
situation persists upon a discontinuous disappearance of the quasiparticle peak, 
leading to the formation of an insulating state with a large Mott gap (shown for $U=3$). 
The formation of heavy quasiparticles can equally be witnessed in the specific heat 
(bottom panel in Fig.~\ref{rhoCv}), with a strong enhancement of the $\gamma=C_V/T$ 
coefficient at the lowest temperature. However, in strong contrast with the usual DMFT
predictions~\cite{DMFTRevModPhys.68.13,SupInfo}, we find the persistence of a finite 
$\gamma$ coefficient in the Mott phase, instead of the usually observed activated behavior 
for a high entropy paramagnetic insulator. This result is in agreement with the thermodynamic
measurements made in several organic materials showing spin liquid behavior.
Fitting from Fig.~\ref{rhoCv} the low-temperature linear slope of $C_V$ in
the Mott insulating phase, we find $\gamma=27k_{B}^2/D$ and $\gamma=14k_{B}^2/D$ 
for $J/D=0.2$ and $J/D=0.4$ respectively. Taking the half-bandwidth in the
range $D=200$ meV, from recent estimates on various organic 
systems~\cite{Powell,KappaBandwidth}, we have approximately $\gamma=80$ mJK$^{-2}$mol$^{-1}$ 
and $\gamma=40$ mJK$^{-2}$mol$^{-1}$ for $J/D=0.2$ and $J/D=0.4$ (these correspond 
to typical values of the exchange constant in organics~\cite{Shimizu}). Our predictions are somewhat 
larger, but of the right magnitude, with the experimental value $\gamma=20$ mJK$^{-2}$mol$^{-1}$ 
measured both for the EtMe$_3$Sb[Pd(dmit)$_2$]$_2$~\cite{SYamashitagapless2011}
and $\kappa$-(BEDT-TTF)$_2$Cu$_2$(CN)$_3$ compounds~\cite{SYamashitagapless}.

We now demonstrate that this thermodynamic effect has a strong influence of the
metal-insulator phase diagram. The main reason is the quenching of the entropy
of the Mott insulator by the presence of a finite exchange interaction $J$, which
typically bends the transition lines towards a stabilization of the metal upon
heating. Indeed, the entropic contribution to free energy of the insulator is 
strongly diminished by exchange, leading to a reentrance of the first order 
transition lines~\cite{Park_Haule_CDMFT,ALiebsch_CDMFT,sergeepl}. 
\begin{figure}[th]
\includegraphics[width=0.99\linewidth,height=0.5\linewidth]{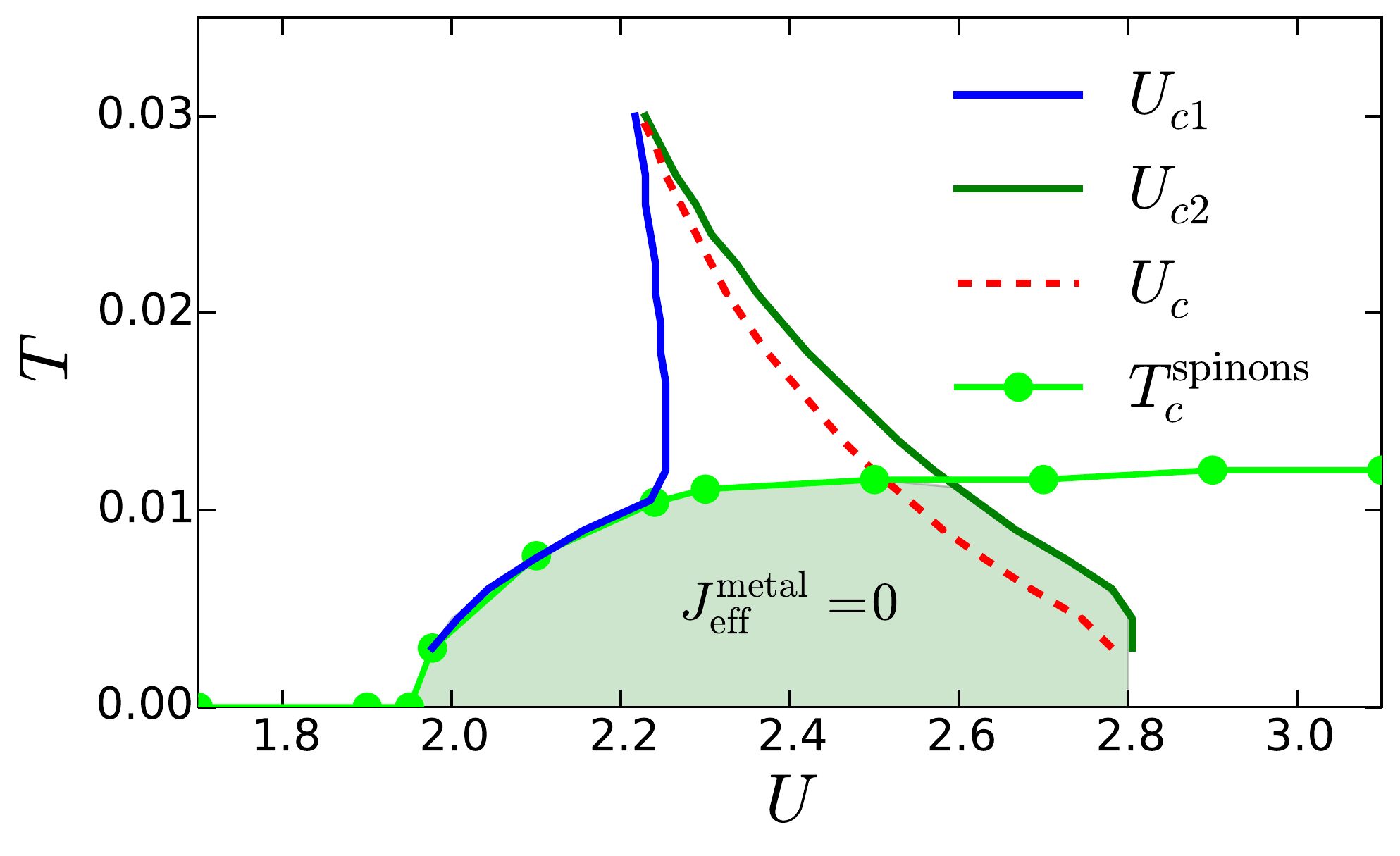}
\includegraphics[width=0.99\linewidth,height=0.5\linewidth]{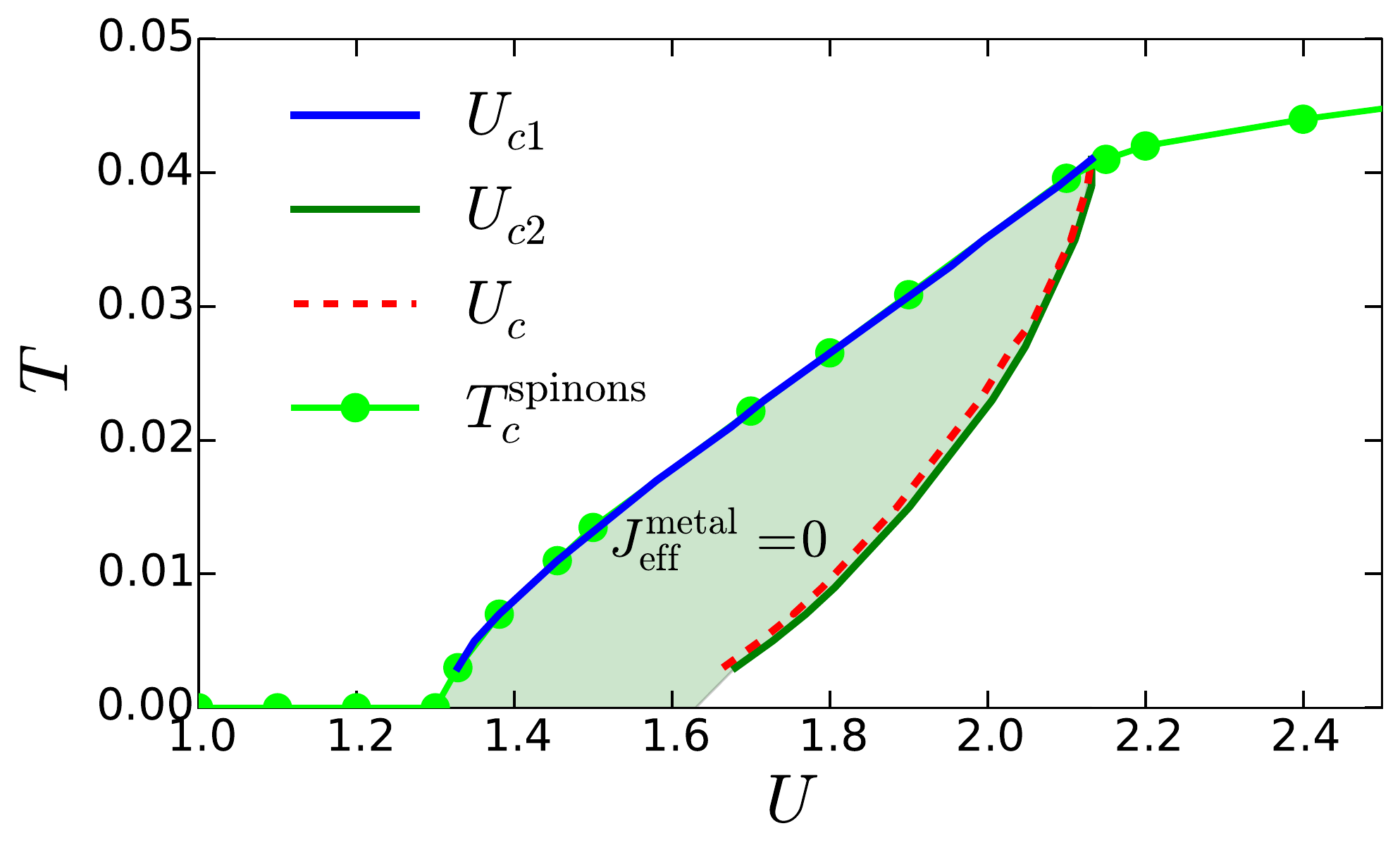}
\caption{(Color online) Phase diagram for $J/D=0.1$ (upper panel) and $J/D=0.4$
(lower panel) as a function of Coulomb strength $U/D$ and temperature $T/D$. 
Continuous lines denote the metal-insulator boundaries, and
dashed lines indicate the true first-order transition based on the free
energy. Also, the region bounded by dots show the low-temperature onset of 
the spinon Fermi surface in the Mott insulator. The shaded region indicates that
$J_\mathrm{eff}^\mathrm{metal}=0$ for the metallic state within the whole coexistence 
region. This demonstrates that spinons are only well defined quasiparticles in 
the low temperature Mott phase, and are never stable in the metal.}
\label{PhaseDiagram}
\end{figure}
We show in Fig.~\ref{PhaseDiagram} two phase diagrams, for $J/D=0.1$ and $J/D=0.4$ 
respectively. The continuous lines denote the metal to insulator boundaries, 
$U_{c1}(T)$ and $U_{c2}(T)$ at which the insulator and metallic solution disappear 
respectively.
For small $J$ (upper panel), only the $U_{c1}$ line is bent, while the whole
transition region is affected for large $J$ (lower panel), with a slight
increase of the maximum critical temperature $T_c$ due to the coupling to
spinon degrees of freedom.
Most importantly, the critical domain moves towards strongly reduced values 
of the Coulomb strength $U$ at increasing $J$, so that the Brinkman-Rice transition,
associated to a diverging $\gamma$ coefficient, is strongly pre-empted by spin
fluctuations. This readily explains the small values of $\gamma$ that were
found in our calculations.

We then consider the fate of the gapless insulating spin liquid. The phase diagrams of 
Fig.~\ref{PhaseDiagram} also show as dots the onset of the spinon Fermi surface,
namely the low-temperature domain where the effective spinon bandwidth
$J_\mathrm{eff}\neq 0$. We find that this domain corresponds 
precisely to the region of existence of the Mott insulator (only at low temperature 
for small $J$, and at temperatures up to the critical $T_c$ of the terminal
Mott endpoint for large enough $J$). Said otherwise, the 
high entropy local moment insulator is always unstable to the formation of a low 
entropy spin liquid, even in the coexistence region of the MIT where 
the Mott gap is strongly reduced. 
Since the gapless spin-liquid state penetrates fully the part 
of the phase diagram where metal and insulator coexist, one can wonder whether the
metallic state is capable of hosting non-trivial spinon excitation. We find
however, for all our simulations, that the spinon bandwidth always vanishes in
the metal, namely $J_\mathrm{eff}^\mathrm{metal}=0$. Since the critical value $U_{c2}(T=0)$ for
the loss of the metal steadily decreases with increasing $J$, this shows that
the Mott transition becomes intrinsically first order at zero temperature, in
contrast to the case $J=0$ (standard DMFT), where the quasiparticle weight 
continuously vanishes.

We finally show that spinons are dramatically repelled from the Fermi
liquid because of the generic occurence of an orthogonality catastrophe in 
the spinon self-energy $\Sigma_f(\w)$ for a Fermi liquid state.
The reason is deeply rooted in the spinon/rotor decomposition, and its
associated constraint $Q = i\partial/\partial \theta_j - 
\sum_\sigma[f_{j\sigma}^{\dagger}f_{j\sigma}-\frac{1}{2}]=0$.
While the spectral function of the physical electron 
$d_{j\sigma}^{\dagger}= f_{j\sigma}^{\dagger} e^{i\theta_j}$
connects excitations within the physical subspace $Q=0$ only, the spinon 
density of states corresponds to processes that link the physical subspace $Q=0$ 
to an unphysical subspace $Q=1$ containing one extra auxiliary slave-particle. 
Thus, the spinon density of states involves matrix elements of the type 
$|\big<\Phi^{(1)}|f^\dagger_{j\sigma}|\Phi^{(0)}\big>|^2$, where $|\Phi^{(Q)}\big>$ are 
wavefunctions living in different $Q$-subspace, which thus experience in a metal
a singular x-ray edge. Due to this mechanism, the spinon self-energy acquires 
an anomalous frequency dependence $\Sigma_f(\w)\propto \w^\alpha$ at low-energy
(typically $\alpha<1/2$).
This crucial property is well obeyed by construction in our impurity 
solver~\cite{sergeprb2_imp_solver}, and can be shown from an exact numerical
solution of quantum impurity problems~\cite{Costi_NCA1,Costi_NCA2}, as well as 
from $1/N$ fluctuations~\cite{ReadFluctuations} around the condensed slave boson 
mean-field theory (upon which the static RVB picture is based). The calculated 
spinon self-energy is shown in Fig.~\ref{Sigmaf} for the metallic and insulating
phases. The anomalously large spinon scattering rate is clearly seen for $U/D=1$,
while a regular behavior is found for $U/D=3$. Spinon are thus very incoherent
in the metallic state, insomuch as to fully destroys their Fermi surface,
as we discuss now.
\begin{figure}[th]
\includegraphics[width=0.9\linewidth,height=0.5\linewidth]{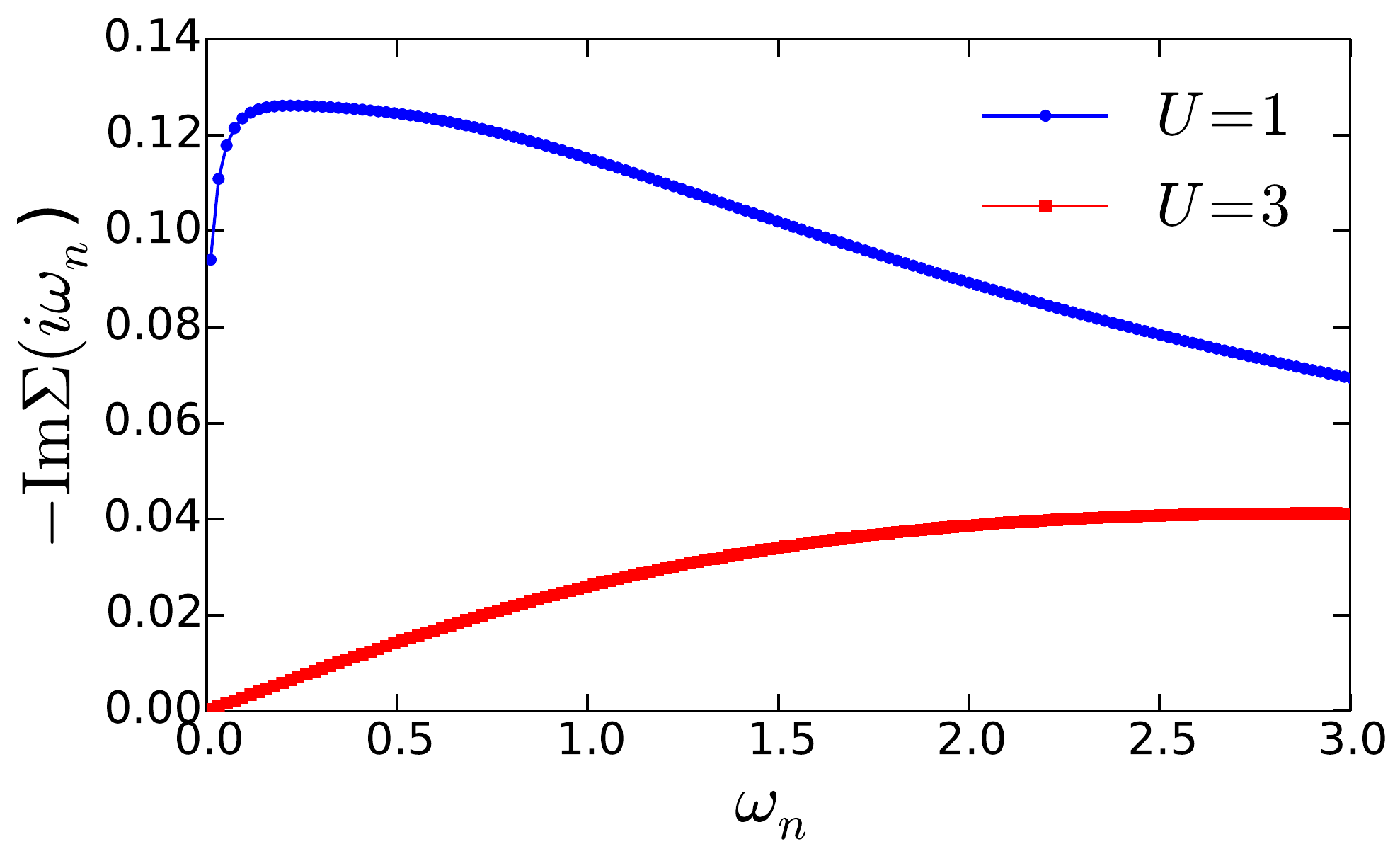}
\caption{(Color online) Frequency-dependent spinon self-energy $\Sigma_f(i\w_n)$
at temperature $T=0.005D$ for $U/D=1$ (metal) and $U/D=3$ (insulator). The
strong enhancement in the metal is due to an x-ray edge effect, which is a
hallmark the spinon spectral density.}
\label{Sigmaf}
\end{figure}

The spinon orthogonality catastrophe is indeed the key physical mechanism
leading to the instability of the spinon Fermi surface in the whole metallic phase. 
The absence of spinons in the metal, $J_\mathrm{eff}^\mathrm{metal}=0$, can be
formulated as a simple inequality from the linearized version of the RVB
equation~(\ref{eq:MF_eq_correction}):
\begin{equation}
\frac{1}{J}>\frac{1}{2\beta}\underset{i\omega_{n}}{\sum}
\frac{1}{[\omega_{n}-\mathrm{Im}\Sigma_{f}(i\omega_{n})]^{2}}.
\label{linearized}
\end{equation}
Using the anomalous spinon self-energy $\Sigma_f(i\w)=-i
C_f|\w|^\alpha\mathrm{Sign}(\w)$ at low energy, and 
evaluating the Matsubara sum at low temperatures, we find
the inequality for stability of the a true Fermi liquid:
\begin{equation}
\frac{1}{J} > \int_0^\infty \frac{\mathrm{d\w}}{\pi}
\frac{1}{[\w+C_f\w^{\alpha}]^2}
\label{inequality}
\end{equation}
Since $\alpha<1/2$, the infrared divergence in the integral is 
cut off and the inequality is fulfilled for small $J$.
In contrast, the Mott insulator is always unstable to the spin-liquid
because the integral diverges for $\Sigma_f=0$.
These analytic arguments are in complete agreement with our numerical 
findings.

In conclusion, we have examined the role of gapless spin excitations,
characterizing frustrated Mott insulators, in the vicinity of the Mott
metal-insulator transition at half-filling. Our result indicates that the
spin-liquid arises simultaneously with Mott gap opening, suggesting that local
magnetic moments generically tend to form a zero-entropy gapless state in
absence of magnetic ordering. However, a spinon Fermi surface cannot be
stabilized upon closing of the Mott gap, due to an orthogonality catastrophe we
identified with emerging charge fluctuations. This mechanism leads to the
conclusion that the Mott transition associated with the loss of quasiparticles
at $U_{c2}$ is inherently of first-order type, even at zero temperature and
for a zero entropy Mott insulator. 
From this perspective, we advocate a scenario 
where behavior consistent with quantum criticality (QC) has purely local
character and emerges only at high temperatures, above the metal-insulator
coexistence region, consistent with experiments. Our theory demonstrates, thus,
that spin liquid correlations do not play a significant role within this high
temperature QC region, which marks the closing of the Mott gap. The physical
picture we propose is dramatically different from the perspective provided from
alternative theories
\cite{sergeprb1_SR_Meanfield,SSLeePatrickLeePRL,SenthilKimSpinliquidPRL,SenthilCriticalFS},
which postulate the dominance of spin-liquid excitation in the entire QC regime
surrounding the Mott point. 

\begin{acknowledgments}
{\it Acknowledgments.} Useful discussions with M. Dressel, S. Fratini, 
K. Kanoda, G. Kotliar, P. Monceaux, I. Paul, A. Ruckenstein and T. Senthil 
are acknowledged. V. D. thanks the CPTGA for supporting a visit to Grenoble. 
Work of T. H. L. and V. D. in Florida was supported by NSF (USA) through 
Grants No. DMR-1005751 and DMR-1410132.
\end{acknowledgments}

\newpage
\setcounter{figure}{0}
\setcounter{table}{0}
\setcounter{equation}{0}

\onecolumngrid

\vspace{1.0cm}
\begin{center}
{\bf Supplemental material: ``Fate of spinons at the Mott point''}
\end{center}

\maketitle

\section{Derivation of the DMFT-RVB equations}

\subsection{Solution in the large $N$ limit}

We consider the Hubbard-Heisenberg Hamiltonian
\begin{equation}
H=-t\underset{<i,j>\sigma}{\sum}d_{i\sigma}^{\dagger}d_{j\sigma}
+U\underset{i}{\sum}\Big[d_{i\uparrow}^{\dagger}d_{i\uparrow}-\frac{1}{2}\Big]
\Big[d_{i\downarrow}^{\dagger}d_{i\downarrow}-\frac{1}{2}\Big]
+J\underset{<i,j>}{\sum}\Big[\underset{\alpha,\alpha'}{\sum}d_{i\alpha}^{\dagger}
\frac{\vec{\tau}_{\alpha,\alpha'}}{2}d_{i\alpha'}\Big]
\cdot\Big[\underset{\sigma,\sigma'}{\sum}d_{j\sigma}^{\dagger}\frac{\vec{\tau}_{\sigma,\sigma'}}{2}d_{j\sigma'}\Big],
\label{Seq:ham}
\end{equation}
with hopping $t$, local Coulomb interaction $U$, and nearest neighbor exchange $J$.
Applying the slave rotor representation, $L_i=-i\frac{\partial}{\partial\theta_{i}}
=\underset{\sigma}{\sum}(f_{i\sigma}^{\dagger}f_{i\sigma}-\frac{1}{2})$
and $d_{i\sigma}=f_{i\sigma}e^{i\theta_{i}}$, the Hamiltonian becomes
\begin{equation}
H=-\frac{U}{2}\underset{i}{\sum}\frac{\partial^{2}}{\partial\theta_{i}^{2}}
-t\underset{<i,j>\sigma}{\sum}f_{i\sigma}^{\dagger}f_{j\sigma}e^{i(\theta_{i}-\theta_{j})}
+J\underset{<i,j>}{\sum}\underset{\sigma,\sigma'}{\sum}\Big[f_{i\sigma}^{\dagger}
\frac{\vec{\tau}_{\sigma,\sigma'}}{2}f_{i\sigma'}\Big]
\cdot\underset{\alpha,\alpha'}{\sum}
\Big[f_{j\alpha}^{\dagger}\frac{\vec{\tau}_{\alpha,\alpha'}}{2}f_{j\alpha'}\Big].
\label{Seq:ham-slave}
\end{equation}

In order to find a solvable extension of Hamiltonian~(\ref{Seq:ham-slave}), we
extend first the spin flavor to SU($N$), taking the spin indices
$\sigma=1,2,\ldots,N$. In the large $N$ limit, and by rescaling the exchange
interaction by a $1/N$ prefactor, the Hamiltonian simplifies into:
\begin{equation}
H=-\frac{U}{2}\underset{i}{\sum}\frac{\partial^{2}}{\partial\theta_{i}^{2}}
-t\underset{<i,j>\sigma}{\sum}f_{i\sigma}^{\dagger}f_{j\sigma}e^{i(\theta_{i}-\theta_{j})}
-\frac{J}{N}\underset{<i,j>}{\sum}\underset{\sigma,\sigma'}{\sum}
f_{i\sigma}^{\dagger}f_{j\sigma}f_{j\sigma'}^{\dagger}f_{i\sigma'}.\label{Seq:ham-largeN}
\end{equation}
This form shows that the RVB mean field becomes exact when $N\to\infty$, with
$J_\mathrm{eff}=\frac{-J}{N}\underset{<i,j>\sigma}{\sum}\big<f_{i\sigma}^{\dagger}f_{j\sigma}\big>$
the effective exchange term. The large-$N$ limit thus leads to a simpler 
effective Hamiltonian, Eq.~(2) of the main text, that reads:
\begin{equation}
H=-\frac{U}{2}\underset{i}{\sum}\frac{\partial^{2}}{\partial\theta_{i}^{2}}
-t\underset{<i,j>\sigma}{\sum}f_{i\sigma}^{\dagger}f_{j\sigma}e^{i(\theta_{i}-\theta_{j})}
+J_\mathrm{eff}\underset{<i,j>\sigma}{\sum}f_{i\sigma}^{\dagger}f_{j\sigma}.\label{Seq:MF_HAM}
\end{equation}

This effective model bears some similarities to a polaronic problem, since the spinons
have aquired a finite bandwidth, proportional to $J_\mathrm{eff}$, and are
coupled via an additional tunneling term, controlled by the bare electronic hopping $t$, to 
a set of bosonic modes, the local rotor fields.
This model can thus be solved exactly in the limit of infinite dimensions using
the DMFT approach. For simplicity, we consider a Bethe lattice, since the
lattice structure becomes irrelevant for large dimensions (however, we keep in
mind that frustration in the original lattice is key to generate the RVB 
spin-liquid state).
The DMFT equations are obtained by scaling appropriately the hopping parameters in 
large $d$, namely $t\rightarrow t/\sqrt{d}$ and $J_\mathrm{eff}\rightarrow
J_\mathrm{eff}/\sqrt{d}$. Using the cavity method, we derive a local effective
action by integrating all modes outside the cavity:
\begin{eqnarray}
\nonumber
S&=&\int_{0}^{\beta}\!\! d\tau\Bigg[\underset{\sigma}{\sum}
f_{\sigma}^{\dagger}(\tau)(\partial_{\tau}-\mu)f_{\sigma}(\tau)
+\frac{U}{2}L^2-iL\partial_\tau \theta \Bigg]
+\int_{0}^{\beta}\!\!d\tau\int_{0}^{\beta}\!\!d\tau'\Delta(\tau-\tau')\underset{\sigma}{\sum}
f_{\sigma}^{\dagger}(\tau)f_{\sigma}(\tau')e^{i[\theta(\tau)-\theta(\tau')]}\\
&&+\int_{0}^{\beta}\!\!d\tau\int_{0}^{\beta}\!\!d\tau'\Delta_{f}(\tau-\tau')
\underset{\sigma}{\sum}f_{\sigma}^{\dagger}(\tau)f_{\sigma}(\tau'),
\label{Seq:local}
\end{eqnarray}
with $\Delta(\tau-\tau')=t^{2}G_{d}(\tau-\tau')$ and
$\Delta_{f}(\tau-\tau')=J_\mathrm{eff}^{2}G_{f}(\tau-\tau')$
the respective hybridization functions of the physical electrons and spinons.

In order to solve the local effective action~(\ref{Seq:local}),
we generalize the rotor from O(2) to O($M$) symmetry, by replacing
$e^{i\theta}\rightarrow X_{\alpha}$,
with the flavor index $\alpha=1,2,...M$ and the constraint 
$\underset{\alpha}{\sum}|X_{\alpha}|^{2}=M$.
Integrating out the angular momentum field $L_\alpha$, the local action becomes:
\begin{eqnarray}
\nonumber
S&=&\int_{0}^{\beta}\!\!d\tau\Bigg[\underset{\sigma}{\sum}f_{\sigma}^{\dagger}(\tau)(\partial_{\tau}-\mu)f_{\sigma}(\tau)
+\underset{\alpha}{\sum}\frac{|\partial_{\tau}X_\alpha(\tau)|^{2}}{2U}
+\lambda(\tau)\underset{\alpha}{\sum}[|X_{\alpha}(\tau)|^{2}-1]\Bigg]
+\int_{0}^{\beta}\!\! d\tau\int_{0}^{\beta}\!\! d\tau'\Delta_{f}(\tau-\tau')\underset{\sigma}{\sum}
f_{\sigma}^{\dagger}(\tau)f_{\sigma}(\tau')\\
&&+\int_{0}^{\beta}\!\! d\tau\int_{0}^{\beta}\!\! d\tau'\frac{\Delta(\tau-\tau')}{M}\underset{\sigma\alpha}{\sum}
f_{\sigma}^{\dagger}(\tau)f_{\sigma}(\tau')X_{\alpha}(\tau)X_{\alpha}^{\dagger}(\tau'),
\label{Seq:action_X}
\end{eqnarray}
with the constraint field $\lambda$ to inforce compactness.
One then introduces bilocal fields $Q(\tau-\tau')$ and $\bar{Q}(\tau-\tau')$,
which are conjugate to $X_{\alpha}(\tau)X_{\alpha}^{\dagger}(\tau')$
and $f_{\sigma}^{\dagger}(\tau)f_{\sigma}(\tau')$ respectively,
to decouple the interacting term:
\begin{eqnarray}
\nonumber
S&=&\int_{0}^{\beta}\!\!d\tau\Bigg[\underset{\sigma}{\sum}f_{\sigma}^{\dagger}(\tau)(\partial_{\tau}-\mu)f_{\sigma}(\tau)
+\underset{\alpha}{\sum}\frac{|\partial_{\tau}X(\tau)|^{2}}{2U}
+\lambda(\tau)\underset{\alpha}{\sum}[|X_{\alpha}(\tau)|^{2}-1]\Bigg]
+\int_{0}^{\beta}\!\!d\tau\int_{0}^{\beta}d\tau'\Delta_{f}(\tau-\tau')\underset{\sigma}{\sum}
f_{\sigma}^{\dagger}(\tau)f_{\sigma}(\tau')\\
&&-\int_{0}^{\beta}\!\!d\tau\int_{0}^{\beta}\!\!d\tau'\Bigg[
\frac{M}{\Delta(\tau-\tau')} \bar{Q}(\tau-\tau') Q(\tau-\tau')
-Q(\tau-\tau')\underset{\alpha}{\sum}X_{\alpha}(\tau)X_{\alpha}^{\dagger}(\tau')
-\bar{Q}(\tau-\tau')\underset{\sigma}{\sum}f_{\sigma}^{\dagger}(\tau)f_{\sigma}(\tau')\Bigg].
\label{Seq:action_bilocal}
\end{eqnarray}
Now this action is controlled by the saddle point of the fields $Q(\tau-\tau')$
and $\bar{Q}(\tau-\tau')$ in the joint large $N$ and large $M$ limit. We define
the Green's functions,
$G_{f}(\tau)=-\big<T_{\tau}f_{\sigma}(\tau)f_{\sigma}^{\dagger}(0)\big>$
and
$G_{X}(\tau)=+\big<T_{\tau}X_{\alpha}(\tau)X_{\alpha}^{\dagger}(0)\big>$.
Applying the saddle point approximation for large $N$ and $M$, we finally obtain
simple self-consistent equations:
\begin{eqnarray}
\label{Seq:Sf}
\Sigma_{f}(\tau)&=&Q(\tau)=\Delta(\tau)G_{X}(\tau),\\
\label{Seq:SX}
\Sigma_{X}(\tau)&=&\bar{Q}(\tau)=\mathcal{N}\Delta(\tau)G_{f}(\tau),\label{Seq:SP_eq}
\end{eqnarray}
with the Matsubara Green's functions,
\begin{eqnarray}
\label{Seq:Gf}
G_{f}(i\omega_{n})^{-1}&=&i\omega_{n}+\mu-\Sigma_{f}(i\omega_{n})-\Delta_{f}(i\omega_{n}),\\
\label{Seq:GX}
G_{X}(i\nu_{n})^{-1}&=&\frac{\nu_{n}^{2}}{U}+\lambda-\Sigma_{X}(i\nu_{n}).\label{Seq:GF}
\end{eqnarray}
and the DMFT self-consistent field:
\begin{equation}
\Delta(\tau) = t^2 G_d(\tau) = t^2 G_f(\tau) G_X(\tau).
\label{Seq:DMFT}
\end{equation}
Note that the parameter $\mathcal{N}=\frac{N}{M}$ is fixed to a constant of
order one, and is typically taken as $\mathcal{N}=3$ in all our calculations
(this choice allows to recover with good approximation the correct DMFT phase
boundaries).
This set of equations can be solved numerically using fast Fourier transforms.
The Green's function and self-energy in real frequency are obtained by analytic continuation 
using Pad\'{e} approximants, leading to the electronic spectral functions shown in the upper 
panel of Fig.~\ref{SrhoCv}.

\subsection{RVB scheme with DMFT feedback}

Our RVB treatment includes the local dynamical scattering between spinons and rotors, 
usually absent in the standard mean field theory, by taking account the local 
spinons self-energy $\Sigma_f$ obtained in DMFT. Since the action of the spinons 
part has the simple form:
\begin{equation}
S_{f}=\int_{0}^{\beta}\!\! d\tau \Bigg[
\underset{i\sigma}{\sum}f_{i\sigma}^{\dagger} \partial_{\tau}f_{i\sigma}
-\frac{J}{N}\underset{<i,j>}{\sum}\underset{\sigma,\sigma'}{\sum}
f_{i\sigma}^{\dagger}f_{j\sigma}f_{j\sigma'}^{\dagger}f_{i\sigma'}\Bigg]
-\int_{0}^{\beta}\!\! d\tau\int_{0}^{\beta}\!\! d\tau' 
\Sigma_f(\tau-\tau') \underset{i\sigma}{\sum}f_{i\sigma}^{\dagger}(\tau) f_{i\sigma}(\tau'),
\label{Seq:action_spinons}
\end{equation}
one obtains a modified RVB equation, 
\[
J_\mathrm{eff}=-\frac{J}{N\beta}\underset{i\omega_{n},\sigma}{\sum}G_{f,\sigma}^{0}[\big<i,j\big>,iz_{n}],\]
with $z_{n}=\omega_{n}-\mathrm{Im}\Sigma_{f}(i\omega_{n})$ and $G_f^{(0)}$ the free
spinon Green's function computed on the underlying lattice.
In the case of the Bethe tree at large $d$, one gets the non-interacting Green's
function between adjacent sites:
\begin{equation}
G_{f}^{0}[\big<i,j\big>,i\omega_{n}]=\frac{-J_\mathrm{eff}}
{\frac{\omega_{n}^{2}}{2}+\omega_{n}\sqrt{\frac{\omega_{n}^{2}}{4}+J_\mathrm{eff}^{2}}+J_\mathrm{eff}^{2}}.
\label{Seq:Gf0}
\end{equation}
so that the modified RVB equation finally reads:
\begin{equation}
J_\mathrm{eff}=\frac{J}{2\beta}\underset{i\omega_{n}}{\sum}\frac{J_\mathrm{eff}}
{\frac{z_{n}^{2}}{2}+\omega_{n}\sqrt{\frac{z_{n}^{2}}{4}+J_\mathrm{eff}^{2}}+J_\mathrm{eff}^{2}}.
\label{Seq:RVB_MF}
\end{equation}
In each DMFT self-consistency loop, we solve the effective impurity model given
by the integral equations~(\ref{Seq:Sf}-\ref{Seq:GX}).
In a second stage, the spinon mean field equation Eq.~(\ref{Seq:RVB_MF}) is subsequently 
solved by a root finding technique. Finally, an outer loop determines the DMFT
hybridization $\Delta(\tau)$ from Eq.~(\ref{Seq:DMFT}), until full convergence is obtained.

\section{Standard DMFT results at $J=0$}
\begin{figure}[th]
\includegraphics[width=0.5\linewidth,height=0.25\linewidth]{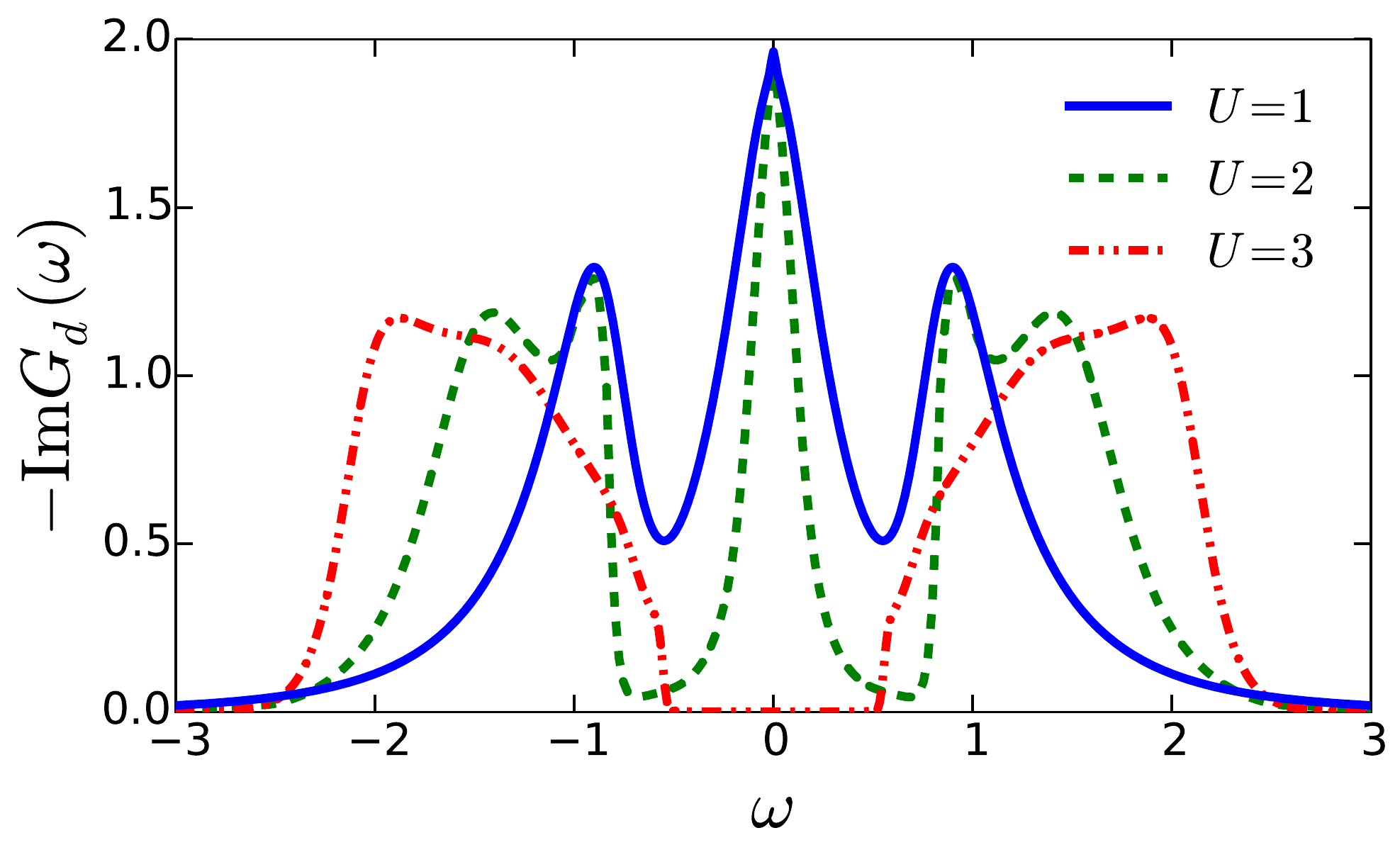}
\vspace{-0.6cm}\\
\includegraphics[width=0.5\linewidth,height=0.25\linewidth]{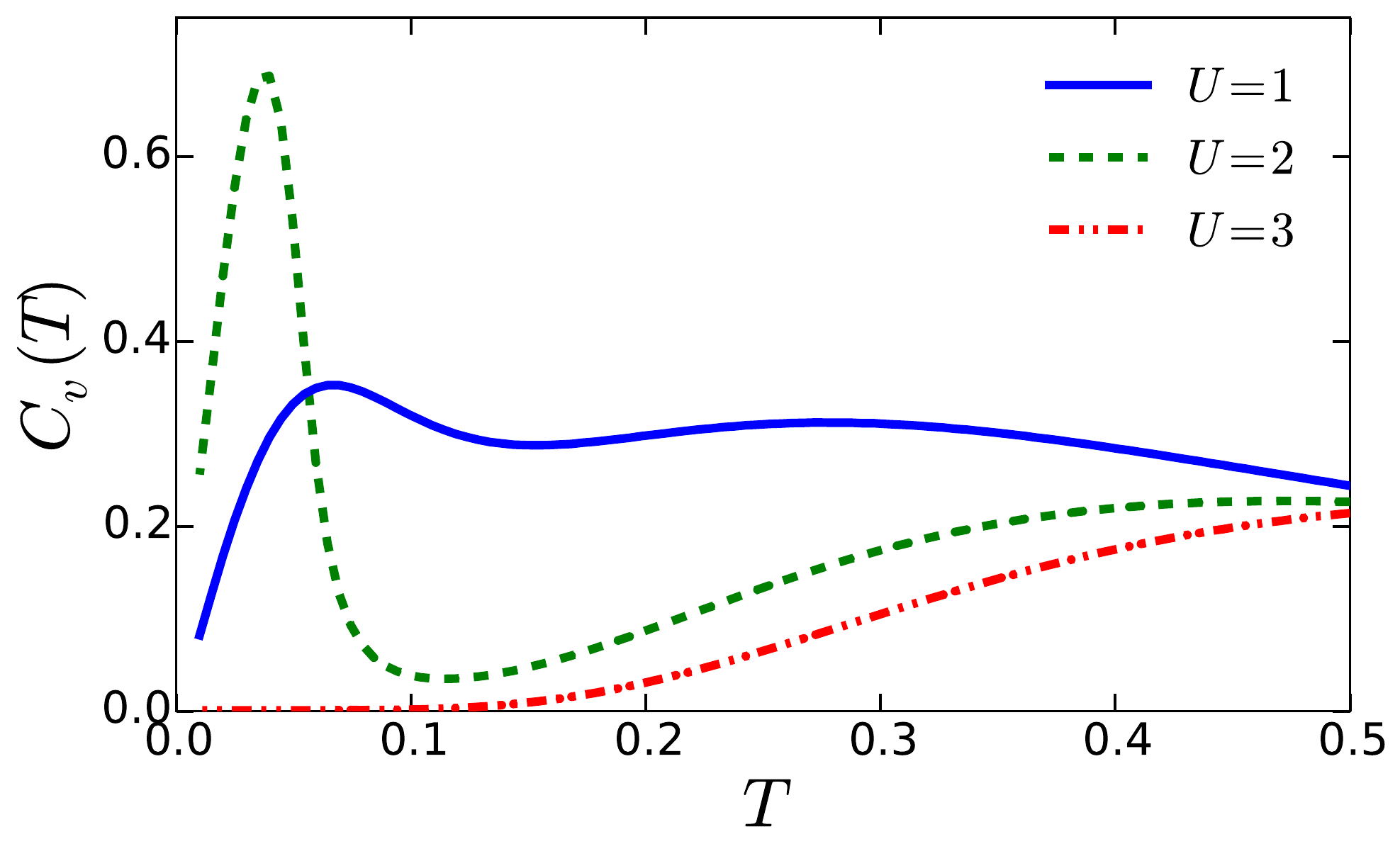}
\caption{Upper panel: electronic density of states across
the Mott transition for $T=0.005D$ and $J/D=0$, with increasing values of the
Coulomb interaction $U/D=1,2,3$. Lower panel: corresponding specific heat
as a function of temperature. The linear specific heat is absent in the Mott 
insulating phase ($U=3$) in the case where antiferromagnetic exchange is turned
off.}
\label{SrhoCv}
\end{figure}

For comparison with the finite $J$ results in the main text, we present here
our calculations for vanishing Heisenberg exchange $J=0$. In this case, spin
fluctuations are suppressed in the Mott insulator, which shows a collection
of local moments associated to a macroscopic entropy.
The density of state in the upper panel of Fig.~\ref{SrhoCv} is remarkably similar 
to the $J=0.2$ results in the main text, for all values of the Coulomb
interaction $U$. In contrast, the linear specific heat
shown in the lower panel of Fig.\ref{SrhoCv} presents a very different behavior
than for finite $J$ in the Mott insulating phase at $U=3$, with an exponential 
suppression of $C_v(T)$ at low temperature, instead of a sharp peak for finite
$J$.

\section{Derivation of the Internal Energy}

Starting from effective Hamiltonian, Eq.~(2) of the main text (adding the
constant mean field contribution, last term in the equation below),
\begin{equation}
H=\underset{\left<i,j\right>\sigma}{\sum}f_{i\sigma}^{\dagger}f_{j\sigma}
[J_\mr{eff}-te^{i(\theta_{i}-\theta_{j})}]
-\frac{U}{2}\underset{i}{\sum}\frac{\partial^{2}}{\partial\theta_{i}^{2}}
+\underset{\left<i,j\right>\sigma}{\sum}\frac{|J_\mr{eff}|^{2}}{J}.
\end{equation}
the internal energy per site (with $N_s$ sites) is calculated by standard 
Green's function methods~\cite{SDMFTRevModPhys.68.13}
\begin{equation}
\label{SInternal}
E_\mathrm{int}=\frac{\big<H\big>}{N_s}
=\frac{T}{N_s}\underset{n,k,\sigma}{\sum}[\epsilon_{k}G_{d\sigma}(k,i\omega_{n})]e^{i\omega_{n}0^{+}}
+\frac{T}{N_s}\underset{n,k,\sigma}{\sum}[\epsilon_{k}G_{f\sigma}(k,i\omega_{n})]e^{i\omega_{n}0^{+}}
+\frac{U}{2}D_{\uparrow\downarrow}+\frac{1}{N_s}\underset{\left<i,j\right>\sigma}{\sum}\frac{|J_\mr{eff}|^{2}}{J}.
\end{equation}
Here, the first and second term correspond to the kinetic energy for electrons
and spinons, that can be computed using a spectral decomposition:
\begin{equation}
\frac{T}{N_s}\underset{n,k,\sigma}{\sum}[\epsilon_{k}G_{d/f,\sigma}(k,i\omega_{n})]e^{i\omega_{n}0^+}
=2T\underset{n}{\sum}e^{i\omega_{n}0^+}
\int d\epsilon\; \epsilon\rho_{d/f}(\epsilon)G_{d/f,\sigma}(\epsilon,i\omega_{n}),
\end{equation}
with the semi-circular density of states of the Bethe lattice for the electronic
density of states, $\rho_{d}(\epsilon)=\frac{1}{\pi t}\sqrt{1-[\epsilon/(2t)]^{2}}$.
The spinons also follow a semi-circular density of states, $\rho_{f}(\epsilon)
=\frac{1}{\pi J_\mr{eff}}\sqrt{1-[\epsilon/(2J_\mr{eff})]^{2}}$, that involves a
different bandwidth $4 J_\mr{eff}$ associated to the spinon dispersion.
Here the lattice electron and spinon Green's functions are given by
$G_{d/f}(\epsilon,i\w_n) =
\frac{1}{i\omega_{n}-\epsilon-\Sigma_{d/f}(i\omega_{n})}$, using the fact that
the DMFT self-energies $\Sigma_{d/f}(i\omega_{n})$ are purely local (but
frequency dependent).

The third term in Eq.~(\ref{SrhoCv}) is associated to the Coulomb interaction,
and is expressed as a function of the double occupancy,
$D_{\uparrow\downarrow}$, which is related to the dynamical charge susceptibility by
$D_{\uparrow\downarrow}=(1/2)\chi_c(\tau=0)$.  The charge susceptibility can be
computed in principle either from spinon response $\chi_c^f$ or from rotor response
$\chi_c^X$\cite{Ssergeprb2_imp_solver}:
\begin{eqnarray}
\nonumber
\chi_c^f(\tau) &=& 
\big<\sum_{\sigma,\sigma'}[f_{j\sigma}^{\dagger}(\tau)f_{j\sigma}(\tau)-\frac{1}{2}]
[f_{j\sigma'}^{\dagger}(0)f_{j\sigma'}(0)-\frac{1}{2}] \big>=2G_{f}(\tau)G_{f}(\tau), \\
\chi_c^X(\tau) &=& \big<i\frac{\partial}{\partial \theta_j}(\tau)  
i\frac{\partial}{\partial \theta_j}(0)\big>
=\frac{2}{U^{2}}\{G_{X}(\tau)[\partial_{\tau}^{2}G_{X}(\tau)+U\delta(\tau)]-[\partial_{\tau}G_{X}(\tau)]^{2}\}.
\end{eqnarray}
Both quantities should be equal in absence of approximation, but they do differ
at the saddle point level. For this reason, we use Nagaosa and Lee composition 
rule~\cite{SSlaveBosonPatrickLeeRevModPhys}, $\chi_c(i\w) =
[(\chi_c^f)^{-1}+(\chi_c^X)^{-1}]^{-1}$,
which allows to recover the correct behavior of the physical charge response
both in the Fermi liquid and in the Mott state.

\end{document}